\documentclass[preprint]{revtex4}
\usepackage{graphicx}
\usepackage{epstopdf}
\usepackage{amsmath}
\usepackage{hyperref}
\usepackage{booktabs}
\usepackage{color}
\usepackage{setspace}

\usepackage{chapterbib}

\begin{document}

%
%
%
%
%
\title{The microscopic pathway to crystallization in supercooled liquids} 

\author{John Russo} 
\affiliation{ {Institute of Industrial Science, University of Tokyo, 4-6-1 Komaba, Meguro-ku, Tokyo 153-8505, Japan} }

\author{Hajime Tanaka\footnote{e-mail address: tanaka@iis.u-tokyo.ac.jp}}
\affiliation{ {Institute of Industrial Science, University of Tokyo, 4-6-1 Komaba, Meguro-ku, Tokyo 153-8505, Japan} }


\begin{abstract}
{\bf
Despite its fundamental and technological importance, a microscopic understanding of the crystallization process is still elusive. By computer simulations of the hard-sphere model we reveal the mechanism by which thermal fluctuations drive the transition from the supercooled liquid state to the crystal state. In particular we show that fluctuations in bond orientational order trigger the nucleation process, contrary to the common belief that the transition is initiated by density fluctuations. Moreover, the analysis of bond orientational fluctuations shows that these not only act as seeds of the nucleation process, but also i) determine the particular polymorph which is to be nucleated from them and ii) at high density favour the formation of fivefold structures which can frustrate the formation of crystals. These results can shed new light on our understanding of the
relationship between crystallization and vitrification.}
\end{abstract}

\maketitle

The liquid-to-solid transition is characterized by the spontaneous breaking of both
positional and orientational symmetry, but how this happens microscopically is still a matter of debate~\cite{kelton2010,
palberg,anderson,AuerR,SearR,GasserR}.
Most approaches, like classical nucleation theory (CNT) or density functional theories (DFT)~\cite{baus,oxtoby},
assume that the crystallization process is primarily controlled by positional 
ordering, with the liquid regarded as a spatially uniform background where nucleation can occur at any location 
with an equal probability.
However experiments~\cite{schoppe,savage,iacopini} and simulations~\cite{malley,kawasaki,schilling,bolhuis} have
recently started to point out deviations from the classical picture of crystallization, suggesting that this
process could be more complex than previously thought.

We argue that for understanding the origin of such deviations it may be crucial to recognize
the role of thermally excited fluctuations in driving the transition from the liquid phase to the crystal phase. 
Fluctuation effects were first identified in globular proteins and colloidal systems close to a metastable critical point,
where crystallization starts with the formation of amorphous high-density aggregates and 
is followed by the actual nucleation event occurring within these fluctuations~\cite{ten1997enhancement,shen,oxtoby2003crystal,lutsko_twostep,lutsko2006theoretical}:
the \emph{two-step} nucleation scenario.  
These studies revealed that the coupling between critical concentration fluctuations and density ordering (crystallization) plays a key role in nucleation. 
Even for a single component liquid, experiments~\cite{schoppe,savage,iacopini} and simulations~\cite{malley,schilling} have 
recently showed the importance of density fluctuations in the initial stage of crystallization, which leads to the formation of precursors. 
Since the \emph{two-step} nucleation scenario looks valid far~\cite{lutsko2006theoretical}
or even in absence~\cite{schilling} of a critical point, it has been suggested that this scenario (in which density fluctuations foreshadow structural ordering) 
could indeed be a general nucleation mechanism.
Independently from the aforementioned two-step scenario, recent simulation works \cite{kawasaki,Kawasaki3D}
have pointed out the importance of another type of fluctuations occurring in the supercooled liquid phase:  
spontaneous critical-like fluctuations of bond orientational order~\cite{tanaka,tanaka_stat}. While the density order parameter
(and in general translational order) is a measure of the relative spacing between the neighbouring particles, bond orientational order expresses instead 
the relative orientation of the (geometrical) bonds between a particle and its neighbouring particles.
In both scenarios, thermal fluctuations promote the formation of crystal precursors, i.e. preordered regions which trigger the nucleation process.
However, since density and bond orientational ordering proceed simultaneously in the process of crystal nucleation,  
it has remained elusive how these order parameters are coupled, and whether any of the two plays a primary role.

In the present work we will investigate precursors in models of colloidal systems in order to elucidate the microscopic
mechanism of crystal formation. 
We use here the word \emph{precursor} as a short term for denoting the region of the liquid's free energy
basin where nucleation is more likely to occur. 
We will first rule out the possibility of a two-step process involving
densification as the first step towards crystallization. We will show instead that the nucleation process
proceeds with the crystalline structures emerging first at liquid-like densities, a process akin to what was reported by some
studies of nucleation in molecular systems~\cite{bagdassarian1994crystal,ten1996numerical}.
By examining the crystallization process in the two dimensional order-parameter space of density and orientational order,
we will show that precursor regions are not characterized by locally denser regions, but by locally bond-oriented
regions, and we will present a novel microscopic explanation of this mechanism.
We will show that these precursor regions not only act as seeds of the nucleation process, but also determine the particular polymorph
which is to be nucleated from them. This new concept implies that polymorphism is already a property of the metastable liquid state.

It is interesting to note that regions of high bond orientational order have also been identified as responsible for
the highly heterogeneous dynamics in deeply supercooled liquids, and could be linked to a growing
structural length at the origin of the glass transition~\cite{ShintaniNP,tanaka}. 
A study of the microscopic properties controlling the crystallization of the liquid is thus of utmost importance
not only in elucidating the pathway to crystallization, but potentially also to explain how crystallization
can be avoided. In this context, we will show that our two-order parameter description provides a
thermodynamic justification of Frank's hypothesis~\cite{frank1952supercooling} that icosahedral clusters of particles
act as inhibitors of crystallization.

In this Article we concentrate on the homogeneous nucleation process for the simplest
nontrivial model of a liquid, hard spheres of diameter $\sigma$, by means
of computer simulations. This system is ideal for studying crystallization
and has already provided tremendous contributions to our basic understanding of crystal nucleation \cite{auer,gasser,zaccarelli}.
In Supplementary Information we extend the generality of our study by applying the same concepts to
very different classes of materials, in particular systems governed by ultrasoft potentials (like polymeric materials)
and tetrahedrally coordinated potentials (like water).

\section*{\large Results}

Let us start by introducing the order parameters used in this study. We will describe here
their basic properties, while for the exact mathematical definition we refer to the Methods section.
We will always adopt a microscopic approach, by studying local order parameters (defined at a particle level).
Since the liquid-to-solid transition is characterized by both translational and
orientational symmetry breaking, we wish to monitor both properties during the crystallization process. 
A good order parameter for translational order, which expresses the relative spacing between particles in the system,
is of course the local density $\rho_i$. This is easily computed by means of Voronoi diagrams, which assign to each particle a
local volume $v_i=1/\rho_i$.
To describe orientational order, which expresses the relative orientation between the neighbours around each particle,
we use the spherical harmonics analysis introduced by
Steinhardt et al.~\cite{steinhardt}. We thus define our bond orientational order parameter as $q_6(i)$,
which is a rotationally invariant scalar defined for each particle $i$. A closely related
order parameter is $Q_6(i)$, which is obtained by coarse-graining $q_6(i)$ over its neighbours.
The importance of $Q_6$ lies in the fact that it is a good order parameter to detect precursor regions, as we will show later.
Finally, to address the question whether crystal nuclei emerge from dense precursors,
we need an order parameter that distinguishes disordered configurations from crystal-like ones. We call this order
parameter $S$ (as for \emph{structure}): for particle $i$ it goes from a value close to $0$ in the liquid-state to a value
close to $12$ (as the number of neighbours in a close-packed structure) in the crystal state.

\subsection*{Composition of crystals during nucleation and growth}

We begin by following $50$ spontaneous crystallization events from the metastable state at reduced pressure $\beta p\sigma^3=17$,
where $\beta=1/k_BT$ and $\sigma$ is the hard-spheres diameter. Under these conditions
nuclei form and dissolve repeatedly, until the appearance of a nucleus which grows over
the critical size and eventually spans the whole system.
For each configuration we identify crystal particles following the criteria
pioneered by Frenkel and co-workers~\cite{auer} (see Methods),
and identify individual clusters via a cluster algorithm.
Figure~\ref{fig:nucleus_growth}a shows
the average number of particles with
local bcc, hcp or fcc coordination within the crystal nuclei, as a function of
their size.
The vertical dashed line in Fig.~\ref{fig:nucleus_growth}a, which indicates the average size of the critical nucleus ($n_{\rm c}\simeq 80$) obtained
from umbrella sampling simulations (see Supplementary Information), separates the nucleation and growth regime. 
Within clusters of size smaller than $n_{\rm c}$,  $(66\pm 1)\%$ of the particles are in local fcc coordination. 
This is markedly different from the ratio for random stacking of hcp and fcc hexagonal planes, $n_{\rm fcc}/n_{\rm hcp}\sim 1$, 
which is predicted from the very small bulk free energy difference (around $0.1\%$ of the thermal energy 
in favour of fcc) between fcc and hcp phases~\cite{bolhuis_entropy,pronk}. This behaviour of hard spheres,
also pointed out in earlier studies including both experiments~\cite{gasser,pusey,versmold,palberg,chaikin}
and simulations~\cite{luchnikov,snook,filion}, remains to our knowledge still unexplained and we will show in the
following a mechanism which accounts for this unbalance.
The inset of Fig.~\ref{fig:nucleus_growth}a shows the average density of the crystalline particles as
a function of the nucleus size. All crystalline phases
form at an average number density of $\sim 1.06\sigma^{-3}$, higher than the metastable liquid density
of $\sim 1.02\sigma^{-3}$. The presence of a jump is of course expected for the averaged order parameters
(both $\rho$ and $q_6$) at a first-order phase transition.
More surprisingly instead, the density at which the smallest crystals start forming is still very far
from the bulk density of the stable crystal ($\rho_s\simeq 1.136\sigma^{-3}$).
Thus the nucleation of the solid phase happens under conditions very far
from the bulk solid. As the crystal grows, both the densities of the fcc and hcp phases gradually increase, whereas 
bcc particles are unable to pack efficiently, and hence do not contribute to the cluster growth. 
Here we note that a bulk bcc crystal is in fact mechanically unstable in hard spheres (meaning that
a bulk bcc crystal will immediately transform into a mixture of fcc and hcp crystals).

Now we turn to the order parameter profiles of crystal nuclei.
Figure~\ref{fig:nucleus_growth}b shows the averaged radial profiles of $\rho(r)$ for
different sizes of the nucleus (indicated by the arrow). The density profiles gradually increase
as the nucleus becomes bigger, but still do not reach the bulk values even for sizes much larger than the critical nucleus size.
This is in stark contrast to the prediction of classical nucleation theory (CNT), according to which critical nuclei share the same thermodynamic properties of the bulk solid phase.
Such deviations from CNT was predicted by non-classical
approaches~\cite{harrowell_oxtoby,bagdassarian1994crystal,baidakov2000comparison,blavette}.
Contrary to a two-step scenario, where densification foreshadows structuring \cite{lutsko2006theoretical,schilling}, 
we find no such an indication, as shown in the inset of Fig.~\ref{fig:nucleus_growth}b,
where the density gap $\Delta\rho$ between the nucleus and the liquid phase is displayed for different radii $R/R_\text{critical}$
(normalized to the value of the critical radius). 
The density of the nucleus grows continuously from the liquid, with an almost linear relationship
between $\Delta\rho$ and the nucleus size $R$. 

Figure~\ref{fig:nucleus_growth}c shows both the density radial profile $\rho(r)$ and the profile of the structural order
parameter $S(r)$ for critical nuclei ($n\sim 80$). Both profiles are normalized as to be unity in the pure \emph{fcc} crystal,
and zero in the bulk liquid phase. Going from the liquid phase ($r=\infty$) to the centre of the nucleus ($r=0$) we see that the
nucleus first develops some structural order at liquid-like densities, and only later does the density increase as well. At the centre of
the nucleus both the structural order parameter and the density are far from their bulk values, but density is lagging behind the
development of structural order. The inset shows the ($S$,$\rho$)-map for nuclei of different sizes. The continuous line is the
classical behaviour, while simulation points always fall in the region of structured precursors, and not locally denser precursors.
We note that the gradual increase of structural order is rather similar to that reported in \cite{bagdassarian1994crystal}, 
where the structural order profile grows both its height and range simultaneously. 
It may be worth noting that the result in Ref. \cite{bagdassarian1994crystal} is derived from a one-order-parameter DFT model, 
where a perfect decoupling of structural order from density 
is implicitly assumed. The introduction of a coupling between density and structural order in the same type of model leads instead
to the saturation of both structural and density order
at the first stages of nucleation~\cite{harrowell_oxtoby,shen}.
This is an interesting point to be studied since, as described later, our results suggest indeed a weak coupling between the two types 
of order parameters. In relation to this, we also note that translational order in DFT is not the same as bond orientational order:   
the former is specific to solid-type fluctuations, but the latter can be linked to both liquid-type and solid-type fluctuations. 

In conclusion we have found no signs of the two-step process involving enrichment at constant size and then growth, contrary to some theoretical predictions~\cite{lutsko2006theoretical,schilling,blavette}.
We rather find that the density increase is foreshadowed by the prestucturing of the nucleus. This lagging 
of densification behind structuring is similar to the results of previous nucleation studies in Lennard-Jones systems~\cite{oxtoby,shen,ten1996numerical,harrowell_oxtoby},
but with the difference that in these studies both density and structural order are already saturated to the equilibrium values 
when the nucleus size slightly exceeds the critical size, whereas not in our case (see Fig. \ref{fig:nucleus_growth}b and c).
Moreover the prestructuring prior to densification has always been ascribed to
the low compressibility of the liquid phase (see, e.g., \cite{oxtoby}). In the next section we will show instead that density fluctuations in the liquid and crystal phase
overlap to a large extent, and that the prestructuring of the nucleus is rather due to the development of orientational order, as
the true first step towards crystallization.

\subsection*{Interplay between density and bond orientational order}

To explain the nucleation pattern unveiled in the previous section, we will address the question of how density ($\rho$) and
orientational order ($q_6$) are coupled.
In Fig.~\ref{fig:maps}a we display ($\rho$,$q_6$)-maps for the metastable liquid (before the appearance of the critical nucleus) at different pressures.
We average separately for particles identified as liquid (liquid branch, dashed line) and crystal (crystal branch, lines with symbols) (see also Fig. S1b in Supplementary Information).
By comparing the relative position of the two branches in the $(\rho,q_6)$-map it is easy to spot the regions of stability of each phase:
the stable branch lies below the metastable branch, having higher orientational order at fixed density
(or conversely, the stable phase can reach the same degree of orientational order at lower packing).
Let us start by examining the system at reduced pressure $\beta p\sigma^3=11$. This pressure is just below the melting
pressure, which is $\beta p\sigma^3=11.54$~\cite{noya2008determination}. As shown in Fig.~\ref{fig:maps}a the liquid branch is always located
below the crystal branch, and it is thus the stable branch for all values of $q_6$ and $\rho$. This result is of course the expected one,
since we are before the melting line. What is surprising is that we are able to determine the relative position of the system
with respect to the melting line by simply looking at its $(\rho,q_6)$ map, without resorting to free energy calculations.
And again as expected, as we increase the pressure a crossover between the two branches appears, with the crystal branch gaining stability.
For clarity we will focus on the curves at $\beta p\sigma^3=17$, which is the same pressure at which we obtained our Fig.~\ref{fig:nucleus_growth}.
At low $\rho$ and $q_6$ the liquid branch is the stable one. The crystal branch remains metastable until it reaches a plateau of
constant $\rho$, where the crossover with the liquid branch occurs. The value of this plateau is $\rho=1.06\sigma^{-3}$ which is exactly
the average density of the onset of crystal formation which we determined in the inset of Fig.~\ref{fig:nucleus_growth}a. After this plateau the crystal
phase thus becomes the stable phase. This means that in the metastable liquid, particles which reach (because of thermal fluctuations) values
of $q_6$ and $\rho$ bigger than the crossover values are in local coordination shells that are transforming from liquid to crystal-like.
The reason why this
process occurs at constant density is clear if we consider the fact that these particles are already embedded
in regions of high orientational order. This means that their neighbours are already highly ordered,
and by means of small local rearrangements
are able to attain the symmetry of the crystal (in practice crossing the threshold which we use to identify crystal particles).
We show an example of such a microscopic process in the snapshots in Fig.~\ref{fig:maps}a, where small local rearrangements (white arrows)
cause a change of the coordination around the central particle from
liquid-like (blue) to crystal-like (red), without changing the local density but by
increasing significantly the orientational order.

If we now follow the curve at higher $q_6$ and $\rho$ a surprising result emerges: a second crossover between the crystal and liquid branches
makes the liquid branch stable again. The density of this crossover is $\rho=1.107\sigma^{-3}$, which corresponds to a volume packing of $\phi\simeq 58\%$
(which is also the conventional value which marks the beginning of the glassy state in hard spheres~\cite{zaccarelli}).
This second crossover tells us that at very high $q_6$ and $\rho$ the crystal phase becomes unfavoured again. Note that these are purely
static results, not affected by the underlying dynamics.
By using bond orientational analysis (see Supplementary Information), the structures responsible for the stability of the liquid branch at high density
are easily identified as particles embedded in icosahedral environments. Icosahedral particles belonging to the liquid branch can attain higher densities than
the corresponding crystal structures, but due to their fivefold symmetry are not able to attain long range translational order.
The second crossover in the $(\rho,q_6)$ map tells us thus that crystals have a stability window, which is limited at low densities by
disordered configurations (larger configurational entropy of liquid particles), and at high density by clusters with icosahedral structure. We have thus shown that icosahedral particles
act as inhibitor to crystallization, as was recently observed in both experiments~\cite{kelton2003first,royall_nature} and simulations~\cite{laso}. 
This is consistent with a scenario that glass-forming ability is controlled by frustration against crystallization, or 
the presence of low free-energy local configurations incompatible with the crystal symmetry in a liquid~\cite{TanakaGJPCM,tanaka_stat,ShintaniNP,taffs2010effect,mathieu_icosahedra}.

We have seen that the crystallization process is driven by the development of orientational order, which explains the
prestructuring of the nuclei at liquid-like densities. Precursor regions are thus easily identified by bond orientational
order alone. The one disadvantage of $q_6$ is that it also reveals the signal from icosahedral environments of particles.
To locate crystal precursors, an effective strategy is to spatially coarse-grain $q_6$~\cite{lechner,kawasaki}, thus enhancing the signal from coherent regions (crystal-like) and suppressing it in disordered
or icosahedral-like regions. This is the order parameter called $Q_6$, which grows continuously from the liquid branch to the crystal branch.
In Fig.~\ref{fig:maps}b we plot, for the metastable liquid (prior to the appearance of the critical nuclei) at pressure $\beta p\sigma^3=17$,
a map in the ($Q_6,\rho)$ plane of the structural order parameter $S$. $S(i)$ quantifies how many
first-shell neighbours of particle $i$ have similar local environments: for a disordered liquid we expect $S$
to be null, whereas for a bulk close-packed crystal to be $12$, i.e. all neighbours share the same environment.
As we can see from Fig.~\ref{fig:maps}b the structural order parameter
grows `continuously' from low $Q_6$ to high $Q_6$ values. Contour lines are almost parallel to the $\rho$ axis, meaning that density
is only weakly coupled to the increase of crystalline structure. In other words, high density regions encompass all possible values
of the $S$, while high $Q_6$ regions are always the most crystalline. So precursor regions are exclusively
controlled by the coarse-grained orientational order parameter, and density fluctuations are not sufficient to
promote crystallization.

\subsection*{Polymorph selection}

\begin{figure}[!t]
 \centering
\includegraphics[width=12cm]{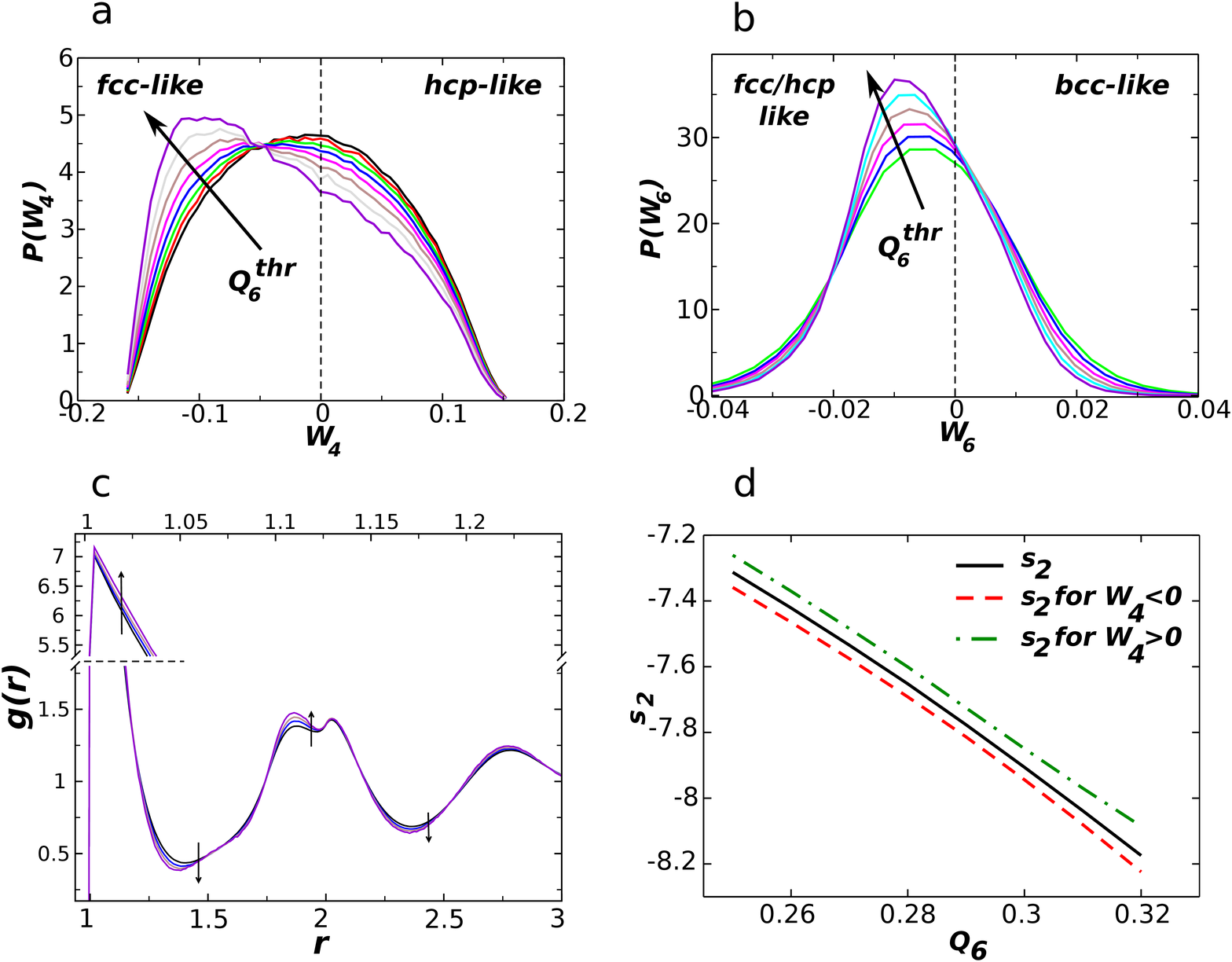}
\caption{{\bf Mechanism of polymorph selection.}
\textbf{a,} Order parameter $W_4$ for liquid particles
having $Q_6>0.25,0.26,0.27,0.28,0.29,0.30,0.31,0.32$ (the order is given by the arrow). The dashed line
is the probability distribution for crystalline particles in the same system. As $Q_6$ increases, the regions
of high structural order in the liquid are characterized by a growing population of fcc-like clusters.
\textbf{b,} Order parameter $W_6$ for liquid particles
having $Q_6>0.27,0.28,0.29,0.30,0.31,0.32$. As $Q_6$ increases, the distributions move to lower
and negative values of $W_6$, thus showing no preference for the bcc symmetry ($W_6>0$).  
Ordering seen in the pair correlation function. 
\textbf{c,} Pair distribution function, $g(r)$, for liquid particles having
$Q_6>0.25,0.28,0.30,0.32$ (the order is given by the arrows). The $y$ axis has been
split to display the first maximum of $g(r)$ (the corresponding $x$ scale is on the top axis).
Regions of high $Q_6$ clearly show an enhanced shoulder in the second peak of the pair
distribution function, which is a precursor to crystallization.
\textbf{d,} Two-body excess entropy $s_2$ (continuous line), calculated for liquid particles with $Q_6>Q_6^\text{thr}$;
the dashed and dotted-dashed lines are instead calculated for liquid particles having
$W_4<0$ and $W_4>0$ respectively. fcc-like particles ($W_4<0$) in regions of high $Q_6$ are thus favoured for
crystallization over hcp-like particle ($W_4>0$).}
 \label{fig:polymorphism}
\end{figure}

Crystals repeatedly appear, grow and melt as represented by the fluctuations in the bond orientational order parameter $Q_6$.
Since crystal nuclei appear from regions
of high bond orientational order, the study of such regions should provide important information on the forming nuclei.
In particular we will show that not only the precursor regions act as seed for crystal growth, but they also determine
which polymorph will be nucleated from them.
To do so we use the order parameters $W_4$ and $W_6$, which are very useful in the detection of polymorphs. We report their
exact definition in the Methods section, and just report here their basic properties. $W_6$ is a good order parameter
to distinguish between bcc crystals and close-packed crystals (hcp and/or fcc), since it is positive in the former whereas 
negative for the latter. $W_4$ is instead good to distinguish between fcc crystals (for which it has negative values) and hcp
crystals (for which it has positive values).
Figure~\ref{fig:polymorphism}a shows the probability distribution for the
order parameter $W_4$ in liquid regions having $Q_6$ higher than a fixed threshold, $Q_6^\text{thr}$.
The $W_4$ distribution was obtained by considering
only liquid particles (crystal particles are not included in the histogram)
in the metastable state (before the critical nucleus is formed),
and the $Q_6^\text{thr}$ threshold values are always within the liquid
distribution. While the metastable liquid has on average a symmetrical distribution around
$W_4=0$, Fig.~\ref{fig:polymorphism}a reveals that the high $Q_6$ regions have a predominant contribution from
negative values of $W_4$, which correspond to the fcc symmetry. Similar histograms are obtained if
instead of thresholds one uses small $Q_6$ intervals centered at progressively high values of $Q_6$ (always within
the liquid distribution).

Since we have shown that crystals form from particles of high $Q_6$, the following scenario emerges for
the nucleation of hard-sphere crystals: the supercooled liquid develops regions of high orientational order (Fig.~\ref{fig:maps}b),
whose symmetry favours the nucleation of the fcc phase (Fig.~\ref{fig:polymorphism}a).
Figure~\ref{fig:polymorphism}b plots the probability distribution for the
order parameter $W_6$, showing that indeed the regions of high $Q_6$ display no preference for the bcc symmetry (characterized by $W_6>0$).
Figure~\ref{fig:polymorphism}c displays the radial distribution function, $g(r)$, for the same high $Q_6$ regions. Notably, higher
$Q_6$ regions show an enhancement of the shoulder in the second peak of the pair distribution function,
which is known to be a structural precursor to the freezing transition~\cite{truskett}. The fact that regions of
high $Q_6$ are more prone to crystallization can also be seen in Fig.~\ref{fig:polymorphism}d,
where the two-body excess entropy~\cite{baranyai,torquato}, $s_2$, is plotted for different values of the threshold $Q_6^\text{thr}$. 
It is known that the two-body excess entropy forms the dominant contribution to the excess entropy,
of the order of $85-90\%$ in simple monoatomic liquids. Its value is $s_2=-6.8$ for the metastable liquid,
and $s_2\cong -10$ for the bulk crystal. The inset shows that the $s_2$ value indeed rapidly decreases for
increasing values of the threshold $Q_6^\text{thr}$. Moreover, the dashed and dotted-dashed lines display
the values of $s_2$ calculated for particles having $W_4<0$ (fcc-like) and $W_4>0$ (hcp-like) respectively,
demonstrating that there is a large difference in the configurational entropy (at the two-particle level) 
between particles having fcc and
hcp symmetry, the former ones being strongly favoured towards crystallization
(the difference between the $s_2$ value of hcp and fcc-like particles is of the order of $1\%$).
This implies that although fcc and hcp have the same free energy in bulk, small clusters of fcc 
symmetry have a lower free energy (lower configurational, but higher correlational entropy) 
than those of hcp symmetry.

\section*{\large Discussions}

In this Article we have studied the process of crystallization from the
perspective of both local translational and bond orientational order.
Crystallization has so far been described by translational ordering of the density field. However, our study 
clearly indicates that symmetry selection due to packing constraint or directional bonding (like in water, see Supplementary Information),
which is represented by bond orientational order, plays a key role in the crystallization process. 
It is bond orientational order and neither density nor translational order 
that triggers crystal nucleation. Structuring before densification was also reported in Refs.~\cite{ten1996numerical,oxtoby2003crystal,oettel2012mode} for 
molecular liquids, in which the range of the interaction is longer than the
size of the constitutive particles. For these liquids, the DFT approach showed that the density of the critical nucleus might deviate significantly 
from the bulk phases, and that the compactification of the nuclei has to be accompanied by an increase of their local structure
(i.e. lattice periodicity). 
This nucleation mechanism is often ascribed to the 
low compressibility of the liquid, which favours structuring prior to densification 
upon nucleation \cite{oxtoby}. 
Contrary to this scenario, our results suggest that this behaviour is a consequence of the weak coupling between density
fluctuations and bond-order fluctuations, with the latter driving the crystallization process.
In other words, it is due to the fact that nuclei form in precursors of high orientational order, where small displacements 
can considerably increase the order with very little density change (see the pictures in Fig. \ref{fig:maps}a). 
The increase in the structural order parameter is thus inherited from a well defined region of the metastable liquid phase space, characterized 
by having high orientational order.
Note that high density regions are not necessarily characterized by
high orientational order, and thus they alone cannot trigger the nucleation process. The existence of these
regions of high density and low orientational order suggests that it is not the low compressibility of the
liquid which is responsible for the coupling between density and orientational order.

Moreover we found that regions of high bond orientational order within the metastable liquid not only act as crystal precursors but
can also determine which particular crystal polymorph will nucleate from them, even when precritical nuclei
(which naturally populate the metastable phase) are disregarded from the analysis.
Since the large population of sub-critical embryos belongs to the same metastable free energy
basin of the liquid (as can be seen for example in Fig.~\ref{fig:maps}b),
it is natural to expect that the emergence of polymorphism will be a continuous process starting in the liquid phase.
Polymorphism develops together with bond orientational order,
highlighting the role of precursor regions in the polymorph selection process.
A liquid has orientationally ordered precursor regions which can exist in a variety of crystal symmetries, according to
some high-dimensional probability distribution. For hard spheres, the projections of this probability distribution along some
reaction coordinates are reported in Fig.~\ref{fig:polymorphism}, and show that
precursor regions with the fcc-symmetry are more
abundant than hcp-symmetry regions. So if a nucleation event occurs in any of these regions, the crystal environment will
reflect the symmetry of the precursor, or the symmetry favoured in a liquid. 
For hard spheres, the preference towards fcc was pointed out in earlier studies, both experiments~\cite{gasser,pusey,versmold,palberg,chaikin}
and simulations~\cite{luchnikov,snook,filion}, and can be explained classically neither by
Ostwald's step rule nor by the Alexander-McTague scenario~\cite{alexander}.
While correctly pointing to the relevance of metastable states, Ostwald's rule cannot be literally
applied to predict the outcome of a nucleation process. 
Instead, we have shown that the relative abundance of one polymorph over the other depends directly 
on the liquid-state precursor's composition. 
This may be related to the scenario proposed by Stranski and Totomanow~\cite{stranski}, where the embryos that form most readily 
are those with the lowest free energy barrier to nucleation. Our results suggest that the physical mechanism behind this rule
is a matching of bond orientational symmetry between precursor regions and crystals, which leads to the reduction of the free barrier for nucleation (the interfacial 
energy). To give a more quantitative account of this scenario,
we also calculated the pair correlation entropy of precursor regions (Fig.~\ref{fig:polymorphism}), showing indeed an imbalance between the
different crystal symmetries. 
We confirm that this scenario of crystallization and the resulting selection mechanism of polymorphs are also 
valid for soft spheres (the Gaussian Core model)~\cite{russo_gcm}
and water (see Supplementary Information). 
These results could help clarifying the mechanism behind the interplay between crystallization and liquid polymorphism
which was recently found for both water~\cite{molinero} and silicon~\cite{desgranges}.

Our two-dimensional analysis also unveiled a density range of stability
of the crystals which continuously form in the metastable liquid. This range of stability is limited
at low densities by the usual disordered liquid configurations, and at high densities by fivefold arrangements of particles.
This result, obtained from purely static arguments, provides a
thermodynamic justification of Frank's hypothesis~\cite{frank1952supercooling} that icosahedral clusters of particles
act as inhibitors of crystallization.
This finding may enhance our understanding of the nature of a supercooled metastable liquid state and 
crystallization,
possibly shedding light on the interplay between crystallization and vitrification~\cite{sanz}.
Liquid and crystal often have very different densities, 
due to the translational order of the latter. However, our study reveals that bond orientational order
is the first step in the pathway from the liquid to the crystal state, and a disconnection of this link by competing orientational orderings 
or random disorder  
may be responsible for the avoidance of crystallization, i.e. vitrification (see Fig. \ref{fig:maps}a and Refs. \cite{ShintaniNP,tanaka_stat}).

\vspace{2cm}

\section*{\large Methods}
We study the crystallization process in a system of $N=4000$ monodisperse hard spheres of diameter $\sigma$ by
means of isothermal-isobaric (NPT) Monte Carlo simulations.
Lengths are given in units of the particle diameter $\sigma$ and
pressure in units of $k_BT/\sigma^3$, where $k_BT=1$.
We place the spheres randomly in a simulation box at packing fraction $\eta=0.5352$ and
equilibrate the system at reduced pressure $\beta p\sigma^3=17.0$. At this
pressure the liquid is metastable with respect to crystallization, with
a difference in chemical potential between the liquid and solid state of
$\beta |\Delta\mu|=0.54$. 
As shown in~\cite{filion} the free energy barrier
between the metastable liquid phase and the crystal phase is $\beta\Delta F\simeq 18$,
and the size of the critical nucleus is $\simeq 80$. Under this conditions crystallization
is a rare event, for which not only long trajectories can be obtained for the supercooled liquid,
but also enough nucleation events can be observed spontaneously.

\begin{figure}[!t]
\centering
\includegraphics[width=11cm]{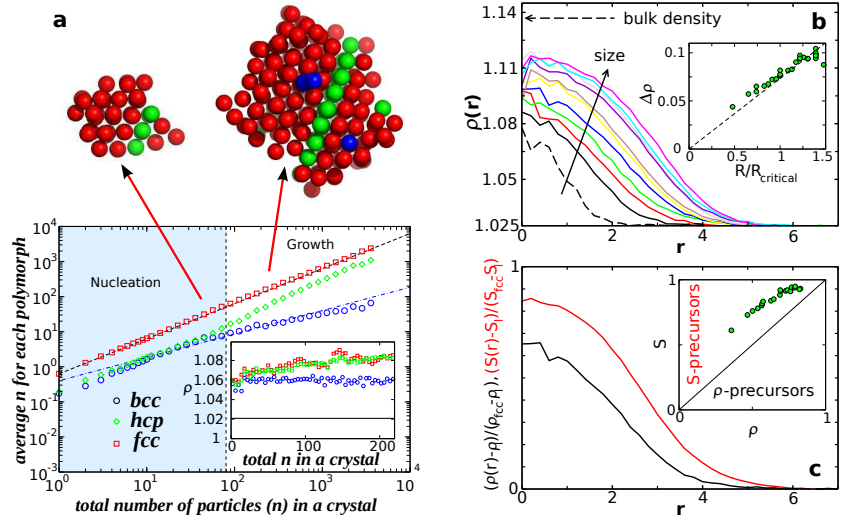}
\caption{
{\bf Composition and radial profiles for crystalline nuclei averaged over many independent trajectories at $\beta p\sigma^3=17$.}
\textbf{a,} Relation between cluster size and polymorphs. 
Average number of particles for bcc (circles), hcp (diamonds) and
fcc (squares) polymorphs as a function of the total crystal size ($n$). The dashed line
grows as the volume, $\sim n$, 
whereas the dashed-dotted line grows as the surface, $\sim n^{2/3}$. 
The vertical dashed line indicates the critical nucleus size $n_{\rm c}$, which separates 
the nucleation regime (the tint blue colour region) and the growth regime.  
The inset shows the average density of particles belonging to the different
polymorphs, and the continuous line the average density of the liquid phase.
Also shown are two
examples of snapshots of crystal nuclei from the computer simulations, at sizes $n=40$ (left) and $n=220$ (right).
The particles are coloured according to the following code: fcc (red), hcp (green), and bcc (blue). 
\textbf{b,} Average density profiles as a function of the distance $r$ from the centre of mass of the nucleus.
Lines are density profiles for nuclei of sizes between $n=5$ and $n=205$
(plotted every $\Delta n=20$ with the order given by the arrow);
each density profile is averaged over nuclei of sizes $n\pm 5$. Crystals are nucleated at conditions very far from the bulk
value, indicated by the dashed horizontal line. The inset shows the density difference $\Delta\rho$ between the centre of the nucleus and
the liquid density, as
a function of the normalized nucleus size ($R/R_\text{critical}$).
\textbf{c,} Comparison between the density profile ($\rho(r)$ black line) and the structural order parameter profile ($S(r)$ red line) for
the critical nucleus (size $n=80$). Both profiles are normalized to be unity in the $fcc$ crystal state, and zero in the liquid phase.
The inset shows the ($S$,$\rho$)-map for nuclei of different sizes (the same as in the panel \emph{b}). The continuous line is the
classical behaviour, while simulation results show that nuclei form in ordered precursors, and not locally denser precursors.
}
 \label{fig:nucleus_growth}
\end{figure}

\begin{figure}[!t]
 \centering
\includegraphics[width=16cm]{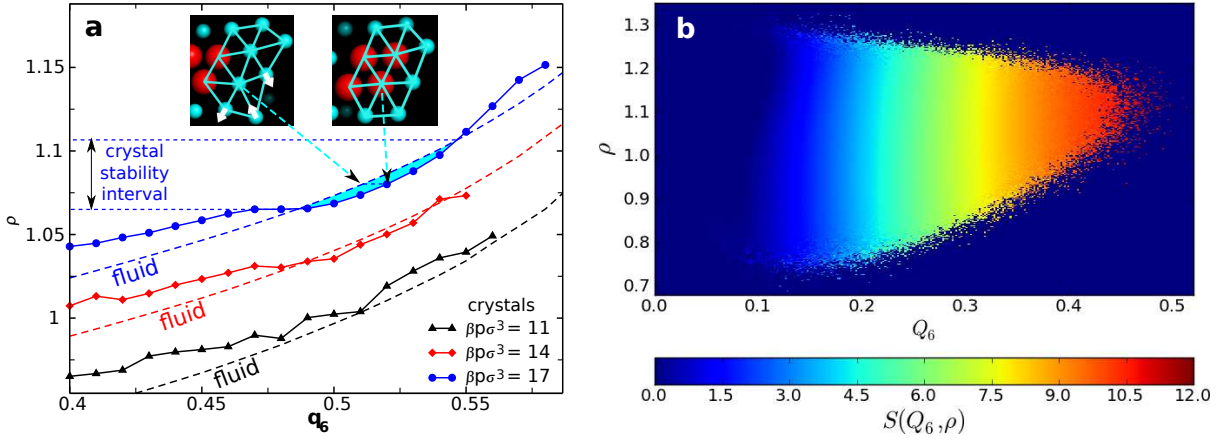}
\caption{
{\bf Roles of density and bond orientational order in crystal nucleation.}
\textbf{a,} Relation between density ($\rho$) and bond orientational order ($q_6$) in the metastable liquid for different pressures, $\beta p\sigma^3=11,14,17$.
Dashed lines are averages over particles in liquid-like environments, whereas full lines+symbols are averages over those in crystal-like environments.
For each pressure, the stable phase is given by the lowest line. For $\beta p\sigma^3=11$, which is just below the melting pressure
$\beta p\sigma^3=11.54$, the liquid line is always stable against the crystal line. But as the pressure is increased the crystal line crosses
the liquid line to become the stable phase. The transition from liquid-like to crystal-like happens at constant density, and can be rationalized
by the small cage rearrangements (as seen in the snapshots) which are sufficient to promote the transition with very little density change. At higher densities a second crossover occurs,
and the liquid branch becomes stable again against the crystal-like branch.
{\bf b,} Probability density for the structural order parameter $S$ in the $(Q_6,\rho)$ plane. The number of connected neighbours
grows continuously from $0$ to $12$ from the liquid to the crystal phase. Contour lines are almost parallel to the $\rho$ axis signalling
that crystallization is promoted mostly by bond orientational order. Regions of high $\rho$ contain particles in a range
of environments from liquid-like to crystal-like, which means that density fluctuations alone are not sufficient to promote crystallization.
}
 \label{fig:maps}
\end{figure}

To identify crystal particles we use the local bond-order analysis introduced by
Steinhardt et al.~\cite{steinhardt}. 
One first introduces a $(2l+1)$ dimensional complex vector ($\mathbf{q}_l$), which
is defined for each particle $i$ as $q_{lm}(i)=\frac{1}{N_b(i)}\sum_{j=1}^{N_b(i)} Y_{lm}(\mathbf{\hat{r}_{ij}})$,
where $l$ is a free integer parameter, $m$ is an integer
that runs from $m=-l$ to $m=l$, $Y_{lm}$ are the spherical harmonics,
$\mathbf{\hat{r}_{ij}}$ is the vector from particle $i$ to particle $j$,
and the sum goes over all neighbouring particles $N_b(i)$ of particle $i$. Since for hard spheres it is known that
the stable crystals are the close packed structures we can impose $N_b(i)=12$,
i.e. we consider only the 12 nearest neighbours (a procedure which is density independent and greatly improves the statistics).
>From the vectors $\mathbf{q}_l$ one can construct different invariants, and our bond orientational order parameter is one of them,
specifically $q_6(i)=\sqrt{4\pi\sum_{m=-6}^{6}|q_{6m}(i)|^2/(2l+1)}$. The vectors $\mathbf{q}_l$ have been proven to be
useful also to identify crystal particles within the liquid. This procedure, first applied to study nucleation by
Frenkel and co-workers~\cite{auer}, consists of comparing the orientational environments of two neighbouring particles
via a scalar product $\mathbf{q}_6(i)/|\mathbf{q}_6(i)|\cdot\mathbf{q}_6(j)/|\mathbf{q}_6(j)|$. If the scalar product between
two neighbours exceeds $0.7$ then the two particles are deemed \emph{connected}. We then identify particle $i$ as being
within a \textit{crystal}
if it is connected with at least $7$ neighbours, and otherwise within a \emph{liquid}.
The structural order parameter, $S(i)$, of a particle $i$ (which we employed in Fig,~\ref{fig:nucleus_growth}c and Fig.~\ref{fig:maps}b)
simply expresses the number of connected neighbours in a continuous way, i.e. $S_i=\sum_{j=1}^{N_b(i)}\frac{\mathbf{q}_6(i)\cdot\mathbf{q}_6(j)}{|\mathbf{q}_6(i)||\mathbf{q}_6(j)|}$.

To distinguish between the different crystal polymorphs we employ the spatially averaged
local bond order parameters introduced in Ref.~\cite{lechner}.
We first define the quantities 
$
\hat{q}_{lm}(i)=\frac{1}{N_b(i)}\sum_{k=0}^{N_b(i)}q_{lm}(k)
$.
Given the previous definition, one can construct the
rotationally invariant quantities
$$
Q_l(i)=\sqrt{4\pi/(2l+1)}|\mathbf{\hat{q}}_l(i)| \nonumber
$$
and
\begin{eqnarray}
W_l(i)=\sum_{m_1,m_2,m_3=0}^l\begin{pmatrix}
l & l & l \\ m_1 & m_2 & m_3
\end{pmatrix} \frac{\hat{q}_{lm_1}(i)\hat{q}_{lm_2}(i)\hat{q}_{lm_3}(i)}{|\mathbf{\hat{q}}_l(i)|^3}  \nonumber
\end{eqnarray}
where the term in parentheses is the Wigner $3-j$ symbol (which is different from
zero only when $m_1+m_2+m_3=0$).

More details about these analyses are given in Supplementary Information.

\vspace{\baselineskip}
\noindent
{\bf Acknowledgments}
The authors thank Mathieu Leocmach, Flavio Romano and Laura Filion for fruitful discussions.
We thank Marjolein Dijkstra, Daan Frenkel and Francesco Sciortino for a critical
reading of an earlier version of the manuscript.
This work was partially supported by a grant-in-aid from the 
Ministry of Education, Culture, Sports, Science and Technology, Japan and 
Aihara Project, the FIRST program from JSPS, initiated by CSTP.

\noindent
{\bf Author Contributions}
H. T. conceived the research, J. R. performed simulations, J. R. and H. T. discussed and
wrote the manuscript.

\noindent
{\bf Additional Information} 
The authors declare that they have no competing financial interests. 
Correspondence and requests for materials should be addressed to 
H. T.

\clearpage

\clearpage

\clearpage

\clearpage



\setlength{\tabcolsep}{10pt}

\newenvironment{sistema}%
  {\left\lbrace\begin{array}{@{}l@{}}}%
  {\end{array}\right.}

\renewcommand{\figurename}{Fig.} 
\renewcommand{\thefigure}{S\arabic{figure}} 



\centerline{\bf \large Supplementary Information for ``The microscopic pathway}
\centerline{\bf \large to crystallization in supercooled liquids''} 
\vspace{0.5cm}

\centerline{John Russo and Hajime Tanaka} 
\centerline{ {\it Institute of Industrial Science, University of Tokyo}}
\centerline{ {\it 4-6-1 Komaba, Meguro-ku, Tokyo 153-8505, Japan} }

\vspace{1cm}

Supplementary Information is organized as follows. First we discuss in more details the methods employed in our study,
with emphasis on the crystal identification protocol. We then show that the microscopic mechanism of crystallization discussed for HS in the main text, also applies to
other relevant classes of systems. In particular we examine:
i) the Gaussian Core model (GCM), a model for self-avoiding polymers in an athermal solvent, 
which belongs to the class of ultrasoft potentials~\cite{stillinger};
ii) the Mw model, a model for water, which belongs to the class of tetrahedrally coordinated liquids~\cite{molinero2008water}.

\vspace{1cm}
\noindent
{\bf Crystal identification}

\begin{figure}[!b]
 \centering
 \includegraphics[width=12cm]{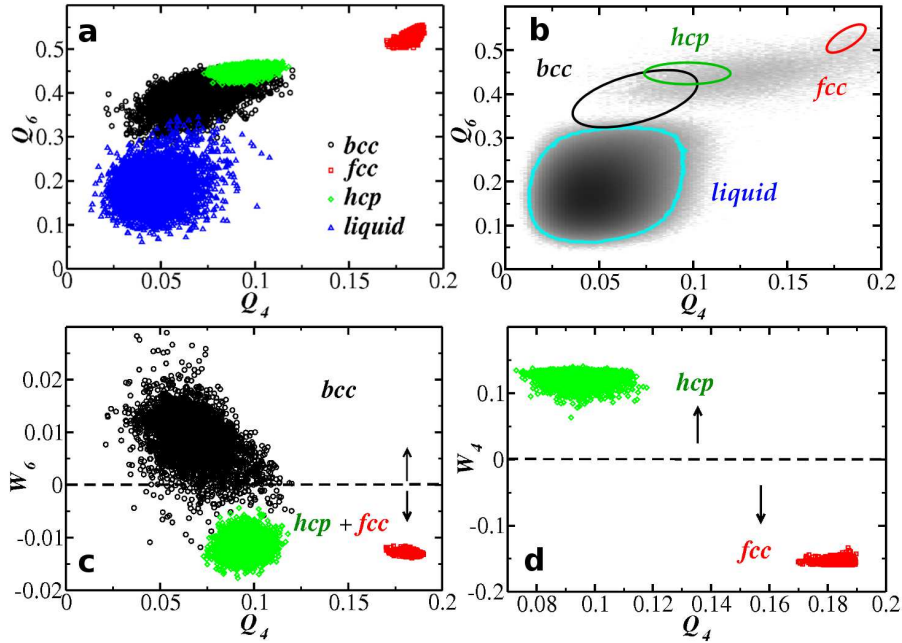}
\caption{{\bf Order parameter maps for the thermal crystals and the supercooled state.}
{\bf a,} $Q_4$-$Q_6$ plane. 
{\bf c,} $Q_4$-$W_6$ plane. {\bf d,} $Q_4$-$W_4$ plane. 
In {\bf b} the probability distribution for
the supercooled state is superimposed on the $Q_4$-$Q_6$ map for the perfect crystals (fcc, hcp and bcc). The maps show that
crystals have higher values of $Q_6$ than the supercooled liquid. 
We can also see that the supercooled liquid prior to crystal nucleation consists of its major liquid portion (the dark gray region; 
$Q_6<0.35$) and minor solid portion  
(the tint gray region; $Q_6>0.35$).
The polymorphs can be identified by exploiting their symmetries along different axes: bcc crystals have $W_6>0$, whereas 
hcp and fcc crystals are both characterized by $W_6<0$ and have respectively $W_4>0$ and $W_4<0$.}
 \label{sfig:maps}
\end{figure}

The definitions of the order parameters employed in our study are reported in the \emph{Methods} section.
Fig.~\ref{sfig:maps} shows the distribution of these order parameters for both the supercooled liquid and
the bulk fcc, hcp and bcc crystals at reduced pressure $\beta p \sigma^3=17$.

The $Q_4$-$Q_6$ map (Fig.~\ref{sfig:maps}a) shows that
crystal structures are always located at higher $Q_6$ than the liquid state. But the map
is not effective for distinguishing between the different polymorphs due to the
large overlap between the bcc and the hcp structures.
Another warning concerning the use of this map for crystal identification regards its
pressure dependence. As we saw in the main text, crystal particles form at conditions very far
from bulk values and in particular the density of formation of the smallest nuclei is much lower
than the final bulk density. By computing the $Q_4-Q_6$ map at different average densities for the
bulk crystals one sees that these maps significantly shift to lower $Q_6$ as density is decreased.
We thus conclude that $Q_4-Q_6$ maps cannot be reliably used for crystal identification.
Instead, to identify the crystal polymorphs we
take advantage of the different symmetries that the crystals
have on the $W_6$ and $W_4$ axis. The bcc structure is in fact characterized by
a positive $W_6$ distribution (Fig.~\ref{sfig:maps}c) whereas hcp and fcc both have negative $W_6$ but
differ respectively for their positive and
negative values of $W_4$ (Fig.~\ref{sfig:maps}d). We have checked that these symmetries are left unchanged
if computing the maps at different average densities (or pressures).
We finally adopt the following criteria for crystal 
classification: First crystal
particles are identified as described in Methods and SI. Then,
we identify i) bcc particles as all crystal particles with $W_6>0$;
ii) hcp particles as all crystal particles with $W_6<0$ and $W_4>0$;
iii) fcc particles as all crystal particles with $W_6<0$ and $W_4<0$.

\begin{figure}[!t]
 \centering
 \includegraphics[width=9cm]{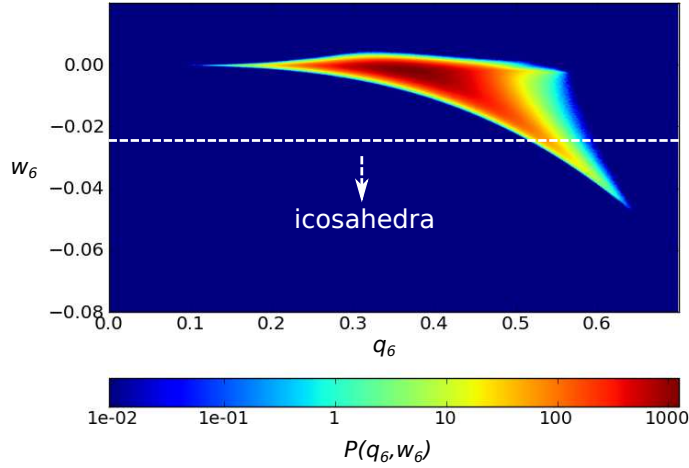}
 \caption{{\bf Probability density map in the $q_6-w_6$ plane for the metastable fluid
at $\beta p\sigma^3=17$.} The map clearly shows that high $q_6$ particles are characterized
by a low value of $w_6$, and are thus particles in icosahedral environment.}
 \label{sfig:q6w6}
\end{figure}

Now we justify the claim made in the article that high $q_6$ particles correspond to
icosahedra. To spot icosahedra we follow the definitions in Ref.~\cite{mathieu_icosahedra},
which we briefly summarize here. Icosahedra can be identified by thresholding the value of
the following order parameter
\begin{eqnarray}
w_6(i)=\sum_{m_1,m_2,m_3=0}^6\begin{pmatrix}
6 & 6 & 6 \\ m_1 & m_2 & m_3
\end{pmatrix} q_{6m_1}(i)q_{6m_2}(i)q_{6m_3}(i)  \nonumber
\end{eqnarray}
where the term in parentheses is the Wigner $3-j$ symbol (which is different from
zero only when $m_1+m_2+m_3=0$), and $q_{lm}$ are the Steinhardt bond orientational
order parameters defined in the Methods section. Following Ref.~\cite{mathieu_icosahedra}
icosahedral particles can be identified as particles having $w_6>0.023$.
As shown in Fig.~\ref{sfig:q6w6}, particles with high $q_6$ all lay within the region which
identifies icosahedral environments. We have thus shown that the high $q_6$ particles, which
form the final liquid stable branch at high densities, are in fact icosahedral particles.

\begin{figure}[!t]
 \centering
 \includegraphics[width=9cm]{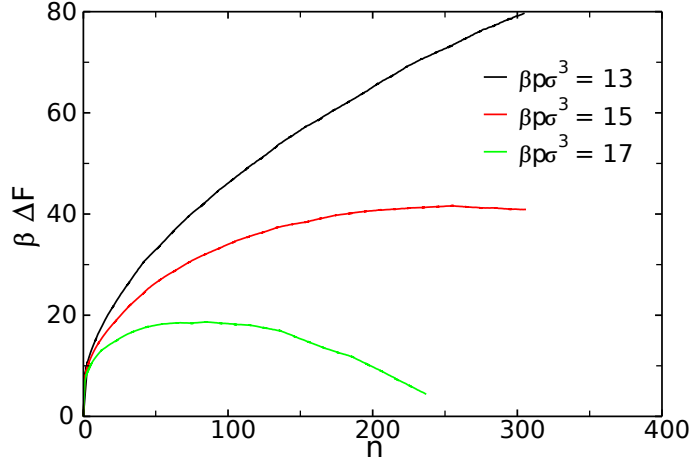}
 \caption{{\bf Free energy barrier for the system at $\beta p\sigma^3=13, 15, 17$, obtained from
Umbrella Sampling simulations.}}
 \label{sfig:barrier}
\end{figure}

With the criteria for identifying crystal particles it is possible to obtain
the free energy barrier and the critical cluster size from
Umbrella Sampling simulations, where a biasing potential is added to
the system Hamiltonian to sample crystalline clusters of large sizes. The details
of the implementation can be found in Ref.~\cite{AuerR}. Fig.~\ref{sfig:barrier} plots
the free energy barrier $\beta\Delta F$ as a function of cluster size.
The free energy barrier
between the metastable liquid phase and the crystal phase at $\beta p\sigma^3=17$ is $\beta\Delta F\simeq 18$,
and the size of the critical nucleus is $\simeq 80$. These results are in good
agreement with the ones in Ref.~\cite{filion}. In this condition crystallization
is a rare event, for which not only long trajectories can be obtained for the supercooled melt,
but also enough nucleation events can be observed spontaneously.

The composition of nuclei obtained from the spontaneously nucleating trajectories were
compared with the ones obtained in equilibrium from the Umbrella Sampling configurations
for nuclei of size up to $250$ particles.
No difference in the average composition of the nuclei was found between the Umbrella Sampling configurations
and the configurations obtained from the Monte Carlo trajectories. This proves that
the small clusters are in quasi-equilibrium, due to the presence of a free energy barrier.

Nuclei composition was calculated also for pressures $\beta p\sigma^3=13,15$ with Umbrella Sampling
configurations and no sensible change in the polymorph composition was found with respect to
the reported pressure $\beta p\sigma^3=17$.

\vspace{1cm}
\noindent
{\bf Gaussian Core model}

The Gaussian Core model (GCM) describes the effective potential between the centres of
mass of polymers dispersed in a good solvent. It consists of pairwise sum of Gaussian components,
first introduced by Stillinger~\cite{stillinger}. The GCM belongs to the class of ultra-soft potentials,
for which there is no divergence at contact. Unlike HS, soft particles can crystallize in
open structures, such as the bcc crystal.
We have recently published a detailed account on the nucleation pathway in the GCM~\cite{russo_gcm}, and thus we
report here the results relevant for our new analysis.
In the following we study the nucleation in the GCM 
for $P=0.05$ and $T=0.0052$, where the units of length and energy are given by the standard deviation and
amplitude of the Gaussian potential (as usual in the literature~\cite{likos}).
According to the phase diagram calculated in Ref.~\cite{pristipino_pre}, the chosen state point has the bcc as
the stable bulk crystal. We follow the crystallization of 200 isobaric Monte Carlo trajectories starting
from a metastable fluid phase. Crystal particles are identified with the following set of
parameters (defined in the previous section), $N_c=9$ and $q_\text{thr}=0.6$, and the different polymorphs
distinguished with the same criteria as HS.

\begin{figure}[!t]
 \centering
 \includegraphics[width=12cm]{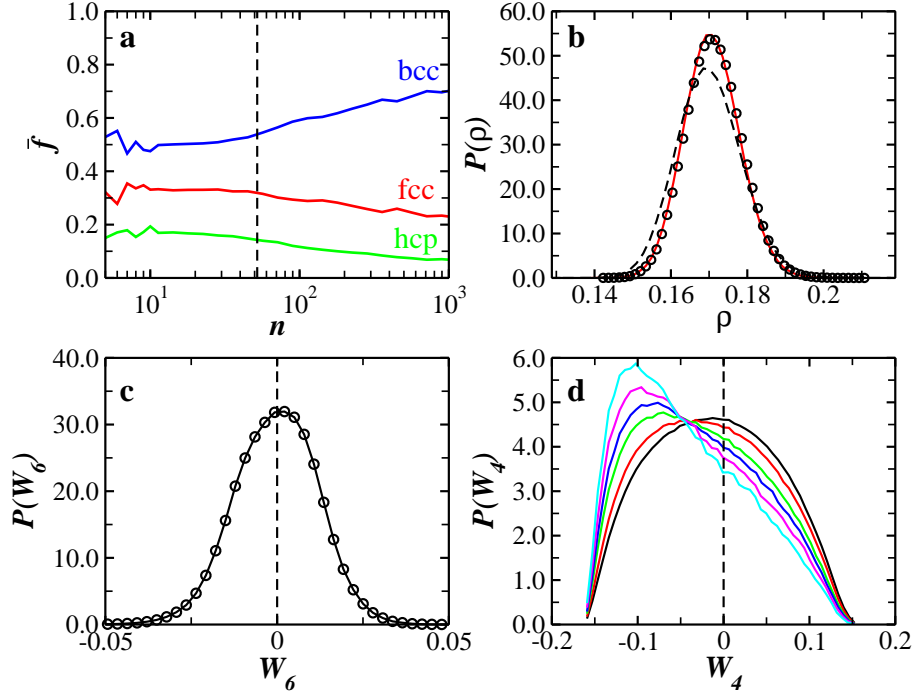}
 \caption{
{\bf Polymorph selection in the GCM.}
 {\bf a,} Fraction of particles ($\bar f$) in a given crystalline state
as a function of the total crystal size $n$ for $P=0.01$. The vertical dashed line denotes the size of the critical nucleus $n_c$.
{\bf b,} Density probability distribution.
The continuous and dashed line are the density histogram for solid and liquid particles respectively. The dots represent the density
histogram for liquid particles fulfilling the condition $Q_6>0.3$.
{\bf c,} $W_4$ probability distribution for liquid particles having $W_6<0$ and $Q_6$ higher than a fixed threshold $Q_6^{\text{thr}}$.
The threshold values plotted are $Q_6>0.25,0.27,0.29,0.30,0.31,0.32$.
{\bf d,} $W_6$ probability distribution for liquid particles having $Q_6>0.3$.}
 \label{sfig:gcm}
\end{figure}

Fig.~\ref{sfig:gcm}a shows the composition of crystalline nuclei as a function of the nucleus size, for
the hcp, fcc and bcc polymorphs. 
The bcc phase is the dominant phase and its fraction increases as the crystal nucleus gets bigger.
Particles in the fcc phase account for $\sim 30\%$ of the
solid particles in the small nuclei, and this fraction decreases as the nuclei become bigger.
The hcp phase accounts only for $\sim 20\%$ of solid particles
in small nuclei, with this fraction steadily decreasing as the nuclei become bigger.
The vertical dashed line in Fig.~\ref{sfig:gcm}a indicates the size of the critical nucleus ($n_c$) obtained
from the mean-first passage time analysis (see Ref.~\cite{wedekind}). The composition of the nucleus for $n<n_c$
is approximately constant, whereas, for $n>n_c$, the fraction of the bcc polymorph
increases at the expenses of both the fcc and hcp phases.

Fig.~\ref{sfig:gcm}b shows the density histogram for liquid (dashed line) and solid particles (continuous line).
Circles in Fig.~\ref{sfig:gcm}b display the density histogram for liquid particles having a value of $Q_6$ higher than $0.3$,
showing that it coincides with the density histogram of the solid particles. Also for the GCM, high $Q_6$ regions are characterized
by the same density fluctuations as the crystalline particles.

From Fig.~\ref{sfig:gcm}a we see that for $n<n_c$ half of the crystalline particles are in the bcc phase, whereas 
the other half are in the fcc or hcp phase. This is consistent with the $W_6$ map shown in Fig.~\ref{sfig:gcm}c,
which shows an almost symmetrical $W_6$ distribution of liquid particles having high $Q_6$. So, unlike HS,
the high $Q_6$ regions in the liquid phase have a symmetry which favours also the bcc phase, promoting its
nucleation. Fig.~\ref{sfig:gcm}a also shows a fraction of fcc particles twice the fraction of hcp
particles. Again this is predicted from the $W_4$ probability distribution function in the liquid phase, depicted
in Fig.~\ref{sfig:gcm}d. As in HS, regions of high orientational order show a preference for the fcc symmetry.

\vspace{1cm}
\noindent
{\bf Mw model for water}

The monoatomic model of water (Mw) is essentially a reparametrization of the Stillinger-Weber potential
to account for the structural and thermodynamic properties of water~\cite{molinero2008water}.
The model has been very successful in describing the supercooled behaviour of water and,
unlike all-atom models, it crystallizes relatively easily.
Because of its distinctive physical properties and its paramount importance, water is a very
good test for our microscopic description of crystallization. Unlike both HS and GCM,
\begin{itemize}
 \item the density of the solid phase is lower than the liquid phase. We should then expect an
anticorrelation between bond orientational order and density, i.e. the density decreasing (instead
of increasing, as in HS) with an increase of bond orientational order;
 \item the solid phases of water at ambient pressure are the hexagonal ice (Ih) and the cubic ice (Ic),
providing a possibility to test our polymorph selection criteria for these target crystals.
\end{itemize}

To detect crystal particles we make use of the CHILL algorithm~\cite{moore2010freezing},
which is a natural extension of the methods described previously for HS to account for
the tetrahedral arrangement of particles. For particle $i$, we define the four closest
neighbours ($j_1\cdots j_4$) and for each bond we compute the scalar product $s_3=\mathbf{\hat{q}}_3(i)\cdot\mathbf{\hat{q}}_3(j)$.
Staggered bonds are characterized by $s_3<-0.8$ and eclipsed bonds by $-0.3<s_3<0.1$.
The Ih crystal is characterized by having one eclipsed bond and three staggered bond, whereas 
Ic crystal has all four bonds in a staggered configuration. Also defective crystal configurations are
detected by following the rules in Ref.~\cite{moore2010freezing}. The definition of these
defects varies somehow in the literature, as in Refs.~\cite{romano_tetrahedral,reinhardt2011free}, but we tested that our results are
independent of the details of this choice.
Testing the symmetry of liquid particles is more subtle. In order to find an order parameter which can distinguish
between the Ic and Ih symmetry, we need to take into account the second coordination shell, which comprises 16 molecules.
In this case we have found that the $W_4$ provides a very good way to distinguish between Ih and Ic, with Ih having
$W_4>0$ and Ic $W_4<0$. 

\begin{figure}[!t]
 \centering
 \includegraphics[width=11cm]{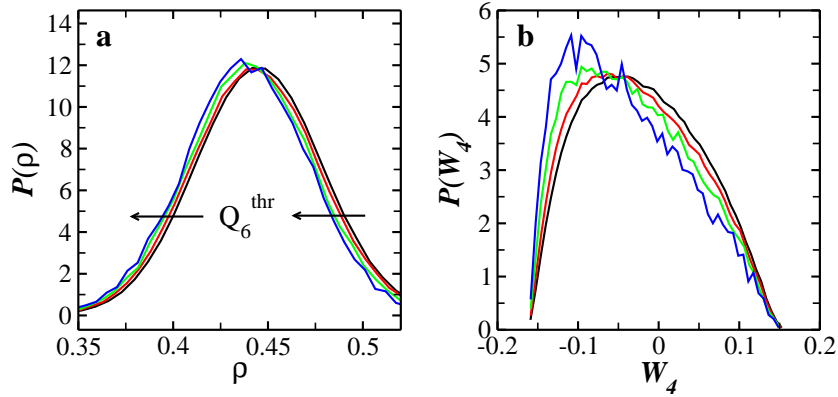}
 
 \caption{ {\bf Polymorph selection in the Mw water.}
{\bf a,} Density probability distribution for liquid particles having $Q_6>Q_6^\text{thr}$,
with $Q_6^\text{thr}=0.09,0.10,0.11,0.12,0.13$ and the order given by the arrow. {\bf b,} $W_4$ probability
distribution for the same set of particles as in panel {\bf a}.}
 \label{sfig:mw}
\end{figure}

Simulations at both $T=180$ K~\cite{moore2010ice,moore2011cubic} and at $T=220$ K~\cite{reinhardt2011free} have shown that
ice spontaneously nucleates preferentially in the Ic form. We run simulations at ambient pressure and
at an intermediate temperature, $T=206$ K,
and confirm the preference for Ic nucleation over Ih nucleation. We then test the symmetry of regions of high
bond orientational order in the liquid phase (not considering crystalline particles).
Fig.~\ref{sfig:mw}a shows the density distribution for liquid particles of high bond orientational order,
having $Q_6>Q_6^\text{thr}$, confirming that indeed regions of high orientational order are anticorrelated
with density. Fig.~\ref{sfig:mw}b finally shows the probability distribution on the $W_4$ axis for liquid particles
with $Q_6>Q_6^\text{thr}$. The distribution becomes more and more peaked towards negative values of $W_4$, which
correspond to the Ic symmetry, as bond orientational order increases in the liquid. This explains why Ic crystals
are the most abundant polymorph at this state point.

We have thus shown that in a large class of potentials (hard, ultrasoft and tetrahedral) nucleation
occurs always in regions of high bond orientational order, and that these regions share the same symmetry
of the nucleating solid phase. This suggests the universality of our scenario.


\begin{thebibliography}{60}
\expandafter\ifx\csname url\endcsname\relax
  \def\url#1{\texttt{#1}}\fi
\expandafter\ifx\csname urlprefix\endcsname\relax\def\urlprefix{URL }\fi
\providecommand{\bibinfo}[2]{#2}
\providecommand{\eprint}[2][]{\url{#2}}

\bibitem{kelton2010}
\bibinfo{author}{Kelton, K.~F.} \& \bibinfo{author}{Greer, A.~L.}
\newblock \emph{\bibinfo{title}{Nucleation in Condensed Matter: Applications in
  Materials and Biology}} (\bibinfo{publisher}{Pergamon},
  \bibinfo{year}{2010}).

\bibitem{palberg}
\bibinfo{author}{Palberg, T.}
\newblock \bibinfo{title}{Colloidal crystallization dynamics}.
\newblock \emph{\bibinfo{journal}{Curr. Opin. Colloid Interface Sci.}}
  \textbf{\bibinfo{volume}{2}}, \bibinfo{pages}{607--614}
  (\bibinfo{year}{1997}).

\bibitem{anderson}
\bibinfo{author}{Anderson, V.~J.} \& \bibinfo{author}{Lekkerkerker, H. N.~W.}
\newblock \bibinfo{title}{Insights into phase transition kinetics from colloid
  science}.
\newblock \emph{\bibinfo{journal}{Nature}} \textbf{\bibinfo{volume}{416}},
  \bibinfo{pages}{811--815} (\bibinfo{year}{2002}).

\bibitem{AuerR}
\bibinfo{author}{Auer, S.} \& \bibinfo{author}{Frenkel, D.}
\newblock \bibinfo{title}{Numerical simulation of crystal nucleation in
  colloids}.
\newblock \emph{\bibinfo{journal}{Adv. Polym. Sci.}}
  \textbf{\bibinfo{volume}{173}}, \bibinfo{pages}{149--207}
  (\bibinfo{year}{2005}).

\bibitem{SearR}
\bibinfo{author}{Sear, R.}
\newblock \bibinfo{title}{{Nucleation: Theory and applications to protein
  solutions and colloidal suspensions}}.
\newblock \emph{\bibinfo{journal}{J. Phys.: Condens. Matter}}
  \textbf{\bibinfo{volume}{19}}, \bibinfo{pages}{033101}
  (\bibinfo{year}{2007}).

\bibitem{GasserR}
\bibinfo{author}{Gasser, U.}
\newblock \bibinfo{title}{Crystallization in three- and two-dimensional
  colloidal suspensions}.
\newblock \emph{\bibinfo{journal}{J. Phys.: Condens. Matter}}
  \textbf{\bibinfo{volume}{21}}, \bibinfo{pages}{203101}
  (\bibinfo{year}{2009}).

\bibitem{baus}
\bibinfo{author}{Baus, M.}
\newblock \bibinfo{title}{Statistical mechanical theories of freezing: An
  overview}.
\newblock \emph{\bibinfo{journal}{J. Stat. Phys.}}
  \textbf{\bibinfo{volume}{48}}, \bibinfo{pages}{1129--1146}
  (\bibinfo{year}{1987}).

\bibitem{oxtoby}
\bibinfo{author}{Oxtoby, D.~W.}
\newblock \bibinfo{title}{Nucleation of first-order phase transitions}.
\newblock \emph{\bibinfo{journal}{Acc. Chem. Res.}}
  \textbf{\bibinfo{volume}{31}}, \bibinfo{pages}{91--97}
  (\bibinfo{year}{1998}).

\bibitem{schoppe}
\bibinfo{author}{Sch\"ope, H.~J.}, \bibinfo{author}{Bryant, G.} \&
  \bibinfo{author}{{van~Megen}, W.}
\newblock \bibinfo{title}{Two-step crystallization kinetics in colloidal
  hard-sphere systems}.
\newblock \emph{\bibinfo{journal}{Phys. Rev. Lett.}}
  \textbf{\bibinfo{volume}{96}}, \bibinfo{pages}{175701}
  (\bibinfo{year}{2006}).

\bibitem{savage}
\bibinfo{author}{Savage, J.~R.} \& \bibinfo{author}{Dinsmore, A.~D.}
\newblock \bibinfo{title}{Experimental evidence for two-step nucleation in
  colloidal crystallization}.
\newblock \emph{\bibinfo{journal}{Phys. Rev. Lett.}}
  \textbf{\bibinfo{volume}{102}}, \bibinfo{pages}{198302}
  (\bibinfo{year}{2009}).

\bibitem{iacopini}
\bibinfo{author}{Iacopini, S.}, \bibinfo{author}{Palberg, T.} \&
  \bibinfo{author}{Sch\"{o}pe, H.~J.}
\newblock \bibinfo{title}{Crystallization kinetics of polydisperse
  hard-sphere-like microgel colloids: Ripening dominated crystal growth above
  melting}.
\newblock \emph{\bibinfo{journal}{J. Chem. Phys.}}
  \textbf{\bibinfo{volume}{130}}, \bibinfo{pages}{084502}
  (\bibinfo{year}{2009}).

\bibitem{malley}
\bibinfo{author}{O'Malley, B.} \& \bibinfo{author}{Snook, I.}
\newblock \bibinfo{title}{Structure of hard-sphere fluid and precursor
  structures to crystallization}.
\newblock \emph{\bibinfo{journal}{J. Chem. Phys.}}
  \textbf{\bibinfo{volume}{123}}, \bibinfo{pages}{054511}
  (\bibinfo{year}{2005}).

\bibitem{kawasaki}
\bibinfo{author}{Kawasaki, T.} \& \bibinfo{author}{Tanaka, H.}
\newblock \bibinfo{title}{{Formation of a crystal nucleus from liquid}}.
\newblock \emph{\bibinfo{journal}{Proc. Nat. Acad. Sci. U.S.A.}}
  \textbf{\bibinfo{volume}{107}}, \bibinfo{pages}{14036}
  (\bibinfo{year}{2010}).

\bibitem{schilling}
\bibinfo{author}{Schilling, T.}, \bibinfo{author}{Sch{\"o}pe, H.~J.},
  \bibinfo{author}{Oettel, M.}, \bibinfo{author}{Opletal, G.} \&
  \bibinfo{author}{Snook, I.}
\newblock \bibinfo{title}{Precursor-mediated crystallization process in
  suspensions of hard spheres}.
\newblock \emph{\bibinfo{journal}{Phys. Rev. Lett.}}
  \textbf{\bibinfo{volume}{105}}, \bibinfo{pages}{25701}
  (\bibinfo{year}{2010}).

\bibitem{bolhuis}
\bibinfo{author}{Lechner, W.}, \bibinfo{author}{Dellago, C.} \&
  \bibinfo{author}{Bolhuis, P.~G.}
\newblock \bibinfo{title}{Role of the prestructured surface cloud in crystal
  nucleation}.
\newblock \emph{\bibinfo{journal}{Phys. Rev. Lett.}}
  \textbf{\bibinfo{volume}{106}}, \bibinfo{pages}{85701}
  (\bibinfo{year}{2011}).

\bibitem{ten1997enhancement}
\bibinfo{author}{{ten Wolde}, P.~R.} \& \bibinfo{author}{Frenkel, D.}
\newblock \bibinfo{title}{Enhancement of protein crystal nucleation by critical
  density fluctuations}.
\newblock \emph{\bibinfo{journal}{Science}} \textbf{\bibinfo{volume}{277}},
  \bibinfo{pages}{1975--1978} (\bibinfo{year}{1997}).

\bibitem{shen}
\bibinfo{author}{Shen, Y.~C.} \& \bibinfo{author}{Oxtoby, D.~W.}
\newblock \bibinfo{title}{Nucleation of Lennard-Jones fluids: A density
  functional approach}.
\newblock \emph{\bibinfo{journal}{J. Chem. Phys.}}
  \textbf{\bibinfo{volume}{105}}, \bibinfo{pages}{6517--6524}
  (\bibinfo{year}{1996}).

\bibitem{oxtoby2003crystal}
\bibinfo{author}{Oxtoby, D.~W.}
\newblock \bibinfo{title}{Crystal nucleation in simple and complex fluids}.
\newblock \emph{\bibinfo{journal}{Philos. T. Roy. Soc. A}}
  \textbf{\bibinfo{volume}{361}}, \bibinfo{pages}{419--428}
  (\bibinfo{year}{2003}).

\bibitem{lutsko_twostep}
\bibinfo{author}{Lutsko, J.~F.}
\newblock \bibinfo{title}{A dynamical theory of nucleation for colloids and
  macromolecules}.
\newblock \emph{\bibinfo{journal}{J. Chem. Phys.}}
  \textbf{\bibinfo{volume}{136}}, \bibinfo{pages}{034509}
  (\bibinfo{year}{2012}).

\bibitem{lutsko2006theoretical}
\bibinfo{author}{Lutsko, J.~F.} \& \bibinfo{author}{Nicolis, G.}
\newblock \bibinfo{title}{Theoretical evidence for a dense fluid precursor to
  crystallization}.
\newblock \emph{\bibinfo{journal}{Phys. Rev. Lett.}}
  \textbf{\bibinfo{volume}{96}}, \bibinfo{pages}{46102} (\bibinfo{year}{2006}).

\bibitem{Kawasaki3D}
\bibinfo{author}{Kawasaki, T.} \& \bibinfo{author}{Tanaka, H.}
\newblock \bibinfo{title}{Structural origin of dynamic heterogeneity in
  three-dimensional colloidal glass formers and its link to crystal
  nucleation}.
\newblock \emph{\bibinfo{journal}{J. Phys.: Condens. Matter}}
  \textbf{\bibinfo{volume}{22}}, \bibinfo{pages}{232102}
  (\bibinfo{year}{2010}).

\bibitem{tanaka}
\bibinfo{author}{Tanaka, H.}, \bibinfo{author}{Kawasaki, T.},
  \bibinfo{author}{Shintani, H.} \& \bibinfo{author}{Watanabe, K.}
\newblock \bibinfo{title}{Critical-like behaviour of glass-forming liquids}.
\newblock \emph{\bibinfo{journal}{Nature Mater.}} \textbf{\bibinfo{volume}{9}},
  \bibinfo{pages}{324--331} (\bibinfo{year}{2010}).

\bibitem{tanaka_stat}
\bibinfo{author}{Tanaka, H.}
\newblock \bibinfo{title}{Bond orientational ordering in a metastable
  supercooled liquid: a shadow of crystallization and liquid-liquid
  transition}.
\newblock \emph{\bibinfo{journal}{J. Stat. Mech.}}
  \textbf{\bibinfo{volume}{2010}}, \bibinfo{pages}{P12001}
  (\bibinfo{year}{2010}).

\bibitem{bagdassarian1994crystal}
\bibinfo{author}{Bagdassarian, C.~K.} \& \bibinfo{author}{Oxtoby, D.~W.}
\newblock \bibinfo{title}{Crystal nucleation and growth from the undercooled
  liquid: A nonclassical piecewise parabolic free-energy model}.
\newblock \emph{\bibinfo{journal}{J. Chem. Phys.}}
  \textbf{\bibinfo{volume}{100}}, \bibinfo{pages}{2139} (\bibinfo{year}{1994}).

\bibitem{ten1996numerical}
\bibinfo{author}{t{en Wolde}, P.~R.}, \bibinfo{author}{Ruiz-Montero, M.~J.} \&
  \bibinfo{author}{Frenkel, D.}
\newblock \bibinfo{title}{Numerical calculation of the rate of crystal
  nucleation in a Lennard-Jones system at moderate undercooling}.
\newblock \emph{\bibinfo{journal}{J. Chem. Phys.}}
  \textbf{\bibinfo{volume}{104}}, \bibinfo{pages}{9932} (\bibinfo{year}{1996}).

\bibitem{ShintaniNP}
\bibinfo{author}{Shintani, H.} \& \bibinfo{author}{Tanaka, H.}
\newblock \bibinfo{title}{Frustration on the way to crystallization in glass}.
\newblock \emph{\bibinfo{journal}{Nature Phys.}} \textbf{\bibinfo{volume}{2}},
  \bibinfo{pages}{200--206} (\bibinfo{year}{2006}).

\bibitem{frank1952supercooling}
\bibinfo{author}{Frank, F.~C.}
\newblock \bibinfo{title}{Supercooling of liquids}.
\newblock \emph{\bibinfo{journal}{P. Roy. Soc. Lond. A Mat.}}
  \textbf{\bibinfo{volume}{215}}, \bibinfo{pages}{43--46}
  (\bibinfo{year}{1952}).

\bibitem{auer}
\bibinfo{author}{Auer, S.} \& \bibinfo{author}{Frenkel, D.}
\newblock \bibinfo{title}{Numerical prediction of absolute crystallization
  rates in hard-sphere colloids}.
\newblock \emph{\bibinfo{journal}{J. Chem. Phys.}}
  \textbf{\bibinfo{volume}{120}}, \bibinfo{pages}{3015--3029}
  (\bibinfo{year}{2004}).

\bibitem{gasser}
\bibinfo{author}{Gasser, U.}, \bibinfo{author}{Weeks, E.~R.},
  \bibinfo{author}{Schofield, A.}, \bibinfo{author}{Pusey, P.~N.} \&
  \bibinfo{author}{Weitz, D.~A.}
\newblock \bibinfo{title}{Real-space imaging of nucleation and growth in
  colloidal crystallization}.
\newblock \emph{\bibinfo{journal}{Science}} \textbf{\bibinfo{volume}{292}},
  \bibinfo{pages}{258} (\bibinfo{year}{2001}).

\bibitem{zaccarelli}
\bibinfo{author}{Zaccarelli, E.} \emph{et~al.}
\newblock \bibinfo{title}{Crystallization of hard-sphere glasses}.
\newblock \emph{\bibinfo{journal}{Phys. Rev. Lett.}}
  \textbf{\bibinfo{volume}{103}}, \bibinfo{pages}{135704}
  (\bibinfo{year}{2009}).

\bibitem{steinhardt}
\bibinfo{author}{Steinhardt, P.~J.}, \bibinfo{author}{Nelson, D.~R.} \&
  \bibinfo{author}{Ronchetti, M.}
\newblock \bibinfo{title}{Bond-orientational order in liquids and glasses}.
\newblock \emph{\bibinfo{journal}{Phys. Rev. B}} \textbf{\bibinfo{volume}{28}},
  \bibinfo{pages}{784--805} (\bibinfo{year}{1983}).

\bibitem{bolhuis_entropy}
\bibinfo{author}{Bolhuis, P.~G.}, \bibinfo{author}{Frenkel, D.},
  \bibinfo{author}{Mau, S.~C.} \& \bibinfo{author}{Huse, D.~A.}
\newblock \bibinfo{title}{Entropy difference between crystal phases}.
\newblock \emph{\bibinfo{journal}{Nature}} \textbf{\bibinfo{volume}{388}},
  \bibinfo{pages}{235--236} (\bibinfo{year}{1997}).

\bibitem{pronk}
\bibinfo{author}{Pronk, S.} \& \bibinfo{author}{Frenkel, D.}
\newblock \bibinfo{title}{Can stacking faults in hard-sphere crystals anneal
  out spontaneously?}
\newblock \emph{\bibinfo{journal}{J. Chem. Phys.}}
  \textbf{\bibinfo{volume}{110}}, \bibinfo{pages}{4589--4592}
  (\bibinfo{year}{1999}).

\bibitem{pusey}
\bibinfo{author}{Pusey, P.~N.} \emph{et~al.}
\newblock \bibinfo{title}{Structure of crystals of hard colloidal spheres}.
\newblock \emph{\bibinfo{journal}{Phys. Rev. Lett.}}
  \textbf{\bibinfo{volume}{63}}, \bibinfo{pages}{2753--2756}
  (\bibinfo{year}{1989}).

\bibitem{versmold}
\bibinfo{author}{Dux, C.} \& \bibinfo{author}{Versmold, H.}
\newblock \bibinfo{title}{Light diffraction from shear ordered colloidal
  dispersions}.
\newblock \emph{\bibinfo{journal}{Phys. Rev. Lett.}}
  \textbf{\bibinfo{volume}{78}}, \bibinfo{pages}{1811--1814}
  (\bibinfo{year}{1997}).

\bibitem{chaikin}
\bibinfo{author}{Cheng, Z.}, \bibinfo{author}{Zhu, J.},
  \bibinfo{author}{Russel, W.~B.}, \bibinfo{author}{Meyer, W.~V.} \&
  \bibinfo{author}{Chaikin, P.~M.}
\newblock \bibinfo{title}{Colloidal hard-sphere crystallization kinetics in
  microgravity and normal gravity}.
\newblock \emph{\bibinfo{journal}{Appl. Optics}} \textbf{\bibinfo{volume}{40}},
  \bibinfo{pages}{4146--4151} (\bibinfo{year}{2001}).

\bibitem{luchnikov}
\bibinfo{author}{Luchnikov, V.}, \bibinfo{author}{Gervois, A.},
  \bibinfo{author}{Richard, P.}, \bibinfo{author}{Oger, L.} \&
  \bibinfo{author}{Troadec, J.~P.}
\newblock \bibinfo{title}{Crystallization of dense hard sphere packings
  Competition of hcp and fcc close order}.
\newblock \emph{\bibinfo{journal}{J. Mol. Liq.}} \textbf{\bibinfo{volume}{96}},
  \bibinfo{pages}{185--194} (\bibinfo{year}{2002}).

\bibitem{snook}
\bibinfo{author}{O'Malley, B.} \& \bibinfo{author}{Snook, I.}
\newblock \bibinfo{title}{Crystal nucleation in the hard sphere system}.
\newblock \emph{\bibinfo{journal}{Phys. Rev. Lett.}}
  \textbf{\bibinfo{volume}{90}}, \bibinfo{pages}{85702} (\bibinfo{year}{2003}).

\bibitem{filion}
\bibinfo{author}{Filion, L.}, \bibinfo{author}{Hermes, M.},
  \bibinfo{author}{Ni, R.} \& \bibinfo{author}{Dijkstra, M.}
\newblock \bibinfo{title}{Crystal nucleation of hard spheres using molecular
  dynamics, umbrella sampling, and forward flux sampling: A comparison of
  simulation techniques}.
\newblock \emph{\bibinfo{journal}{J. Chem. Phys.}}
  \textbf{\bibinfo{volume}{133}}, \bibinfo{pages}{244115}
  (\bibinfo{year}{2010}).

\bibitem{harrowell_oxtoby}
\bibinfo{author}{Harrowell, P.} \& \bibinfo{author}{Oxtoby, D.~W.}
\newblock \bibinfo{title}{A molecular theory of crystal nucleation from the
  melt}.
\newblock \emph{\bibinfo{journal}{J. Chem. Phys.}}
  \textbf{\bibinfo{volume}{80}}, \bibinfo{pages}{1639--1646}
  (\bibinfo{year}{1984}).

\bibitem{baidakov2000comparison}
\bibinfo{author}{Baidakov, V.}, \bibinfo{author}{Boltashev, G.} \&
  \bibinfo{author}{Schmelzer, J.}
\newblock \bibinfo{title}{Comparison of different approaches to the
  determination of the work of critical cluster formation}.
\newblock \emph{\bibinfo{journal}{J. Colloid Interf. Sci.}}
  \textbf{\bibinfo{volume}{231}}, \bibinfo{pages}{312--321}
  (\bibinfo{year}{2000}).

\bibitem{blavette}
\bibinfo{author}{Philippe, T.} \& \bibinfo{author}{Blavette, D.}
\newblock \bibinfo{title}{Minimum free-energy pathway of nucleation}.
\newblock \emph{\bibinfo{journal}{J. Chem. Phys.}}
  \textbf{\bibinfo{volume}{135}}, \bibinfo{pages}{134508}
  (\bibinfo{year}{2011}).

\bibitem{noya2008determination}
\bibinfo{author}{Noya, E.~G.}, \bibinfo{author}{Vega, C.} \&
  \bibinfo{author}{{de Miguel}, E.}
\newblock \bibinfo{title}{Determination of the melting point of hard spheres
  from direct coexistence simulation methods}.
\newblock \emph{\bibinfo{journal}{J. Chem. Phys.}}
  \textbf{\bibinfo{volume}{128}}, \bibinfo{pages}{154507}
  (\bibinfo{year}{2008}).

\bibitem{kelton2003first}
\bibinfo{author}{Kelton, K.~F.} \emph{et~al.}
\newblock \bibinfo{title}{First X-ray scattering studies on electrostatically
  levitated metallic liquids: demonstrated influence of local icosahedral order
  on the nucleation barrier}.
\newblock \emph{\bibinfo{journal}{Phys. Rev. Lett.}}
  \textbf{\bibinfo{volume}{90}}, \bibinfo{pages}{195504}
  (\bibinfo{year}{2003}).

\bibitem{royall_nature}
\bibinfo{author}{Royall, C.~P.}, \bibinfo{author}{Williams, S.~R.},
  \bibinfo{author}{Ohtsuka, T.} \& \bibinfo{author}{Tanaka, H.}
\newblock \bibinfo{title}{Direct observation of a local structural mechanism
  for dynamic arrest}.
\newblock \emph{\bibinfo{journal}{Nature Mater.}} \textbf{\bibinfo{volume}{7}},
  \bibinfo{pages}{556--561} (\bibinfo{year}{2008}).

\bibitem{laso}
\bibinfo{author}{Karayiannis, N.~C.}, \bibinfo{author}{Malshe, R.},
  \bibinfo{author}{de~Pablo, J.~J.} \& \bibinfo{author}{Laso, M.}
\newblock \bibinfo{title}{Fivefold symmetry as an inhibitor to hard-sphere
  crystallization}.
\newblock \emph{\bibinfo{journal}{Phys. Rev. E}} \textbf{\bibinfo{volume}{83}},
  \bibinfo{pages}{061505} (\bibinfo{year}{2011}).

\bibitem{TanakaGJPCM}
\bibinfo{author}{Tanaka, H.}
\newblock \bibinfo{title}{A simple physical model of liquid-glass transition:
  Intrinsic fluctuating interactions and random fields hidden in glass-forming
  liquids}.
\newblock \emph{\bibinfo{journal}{J. Phys.: Condens. Matter}}
  \textbf{\bibinfo{volume}{10}}, \bibinfo{pages}{L207--L214}
  (\bibinfo{year}{1998}).

\bibitem{taffs2010effect}
\bibinfo{author}{Taffs, J.}, \bibinfo{author}{Malins, A.},
  \bibinfo{author}{Williams, S.} \& \bibinfo{author}{Royall, C.}
\newblock \bibinfo{title}{The effect of attractions on the local structure of
  liquids and colloidal fluids}.
\newblock \emph{\bibinfo{journal}{J. Chem. Phys.}}
  \textbf{\bibinfo{volume}{133}}, \bibinfo{pages}{244901}
  (\bibinfo{year}{2010}).

\bibitem{mathieu_icosahedra}
\bibinfo{author}{Leocmach, M.} \& \bibinfo{author}{Tanaka, H.}
\newblock \bibinfo{title}{Roles of icosahedral and crystal-like order in hard
  spheres glass transition}
\newblock \emph{\bibinfo{journal}{Nat. Commun.}}
  \bibinfo{pages}{in print} (\bibinfo{year}{2012}).

\bibitem{lechner}
\bibinfo{author}{Lechner, W.} \& \bibinfo{author}{Dellago, C.}
\newblock \bibinfo{title}{Accurate determination of crystal structures based on
  averaged local bond order parameters}.
\newblock \emph{\bibinfo{journal}{J. Chem. Phys.}}
  \textbf{\bibinfo{volume}{129}}, \bibinfo{pages}{114707}
  (\bibinfo{year}{2008}).

\bibitem{truskett}
\bibinfo{author}{Truskett, T.~M.}, \bibinfo{author}{Torquato, S.},
  \bibinfo{author}{Sastry, S.}, \bibinfo{author}{Debenedetti, P.~G.} \&
  \bibinfo{author}{Stillinger, F.~H.}
\newblock \bibinfo{title}{Structural precursor to freezing in the hard-disk and
  hard-sphere systems}.
\newblock \emph{\bibinfo{journal}{Phys. Rev. E}} \textbf{\bibinfo{volume}{58}},
  \bibinfo{pages}{3083} (\bibinfo{year}{1998}).

\bibitem{baranyai}
\bibinfo{author}{Baranyai, A.} \& \bibinfo{author}{Evans, D.~J.}
\newblock \bibinfo{title}{Direct entropy calculation from computer simulation
  of liquids}.
\newblock \emph{\bibinfo{journal}{Phys. Rev. A}} \textbf{\bibinfo{volume}{40}},
  \bibinfo{pages}{3817} (\bibinfo{year}{1989}).

\bibitem{torquato}
\bibinfo{author}{Truskett, T.~M.}, \bibinfo{author}{Torquato, S.} \&
  \bibinfo{author}{Debenedetti, P.~G.}
\newblock \bibinfo{title}{Towards a quantification of disorder in materials:
  Distinguishing equilibrium and glassy sphere packings}.
\newblock \emph{\bibinfo{journal}{Phys. Rev. E}} \textbf{\bibinfo{volume}{62}},
  \bibinfo{pages}{993} (\bibinfo{year}{2000}).

\bibitem{oettel2012mode}
\bibinfo{author}{Oettel, M.}
\newblock \bibinfo{title}{Mode expansion for the density profile of
  crystal-fluid interfaces: Hard spheres as a test case}.
\newblock \emph{\bibinfo{journal}{arXiv:1203.3756}}  (\bibinfo{year}{2012}).

\bibitem{alexander}
\bibinfo{author}{Alexander, S.} \& \bibinfo{author}{McTague, J.}
\newblock \bibinfo{title}{Should all crystals be bcc? Landau theory of
  solidification and crystal nucleation}.
\newblock \emph{\bibinfo{journal}{Phys. Rev. Lett.}}
  \textbf{\bibinfo{volume}{41}}, \bibinfo{pages}{702--705}
  (\bibinfo{year}{1978}).

\bibitem{stranski}
\bibinfo{author}{Stranski, N.~I.} \& \bibinfo{author}{Totomanow, D.}
\newblock \bibinfo{title}{Rate of formation of (crystal) nuclei and the Ostwald
  step rule.}
\newblock \emph{\bibinfo{journal}{Z. Phys. Chem.}}
  \textbf{\bibinfo{volume}{163}}, \bibinfo{pages}{399--408}
  (\bibinfo{year}{1933}).

\bibitem{russo_gcm}
\bibinfo{author}{Russo, J.} \& \bibinfo{author}{Tanaka, H.}
\newblock \bibinfo{title}{Selection mechanism of polymorphs in the crystal
  nucleation of the Gaussian core model}.
\newblock \emph{\bibinfo{journal}{Soft Matter}} \textbf{\bibinfo{volume}{8}},
  \bibinfo{pages}{4206--4215} (\bibinfo{year}{2012}).

\bibitem{molinero}
\bibinfo{author}{More, E.~B.} \& \bibinfo{author}{Molinero, V.}
\newblock \bibinfo{title}{Structural transformation in supercooled water
  controls the crystallization rate of ice}.
\newblock \emph{\bibinfo{journal}{arXiv:1107.1622v1}}  (\bibinfo{year}{2011}).

\bibitem{desgranges}
\bibinfo{author}{Desgranges, C.} \& \bibinfo{author}{Delhommelle, J.}
\newblock \bibinfo{title}{Role of liquid polymorphism during the
  crystallization of Silicon}.
\newblock \emph{\bibinfo{journal}{J. Am. Chem. Soc.}}  (\bibinfo{year}{2011}).

\bibitem{sanz}
\bibinfo{author}{Sanz, E.} \emph{et~al.}
\newblock \bibinfo{title}{Crystallization mechanism of hard sphere glasses}.
\newblock \emph{\bibinfo{journal}{Phys. Rev. Lett.}}
  \textbf{\bibinfo{volume}{106}}, \bibinfo{pages}{215701}
  (\bibinfo{year}{2011}).

\end{thebibliography}

\begin{thebibliography}{10}
\expandafter\ifx\csname url\endcsname\relax
  \def\url#1{\texttt{#1}}\fi
\expandafter\ifx\csname urlprefix\endcsname\relax\def\urlprefix{URL }\fi
\providecommand{\bibinfo}[2]{#2}
\providecommand{\eprint}[2][]{\url{#2}}

\bibitem{stillinger}
\bibinfo{author}{Stillinger, F.~H.}
\newblock \bibinfo{title}{Phase transitions in the Gaussian core system}.
\newblock \emph{\bibinfo{journal}{J. Chem. Phys.}}
  \textbf{\bibinfo{volume}{65}}, \bibinfo{pages}{3968--3974}
  (\bibinfo{year}{1976}).

\bibitem{molinero2008water}
\bibinfo{author}{Molinero, V.} \& \bibinfo{author}{Moore, E.~B.}
\newblock \bibinfo{title}{Water modeled as an intermediate element between
  Carbon and Silicon }.
\newblock \emph{\bibinfo{journal}{J. Phys. Chem. B}}
  \textbf{\bibinfo{volume}{113}}, \bibinfo{pages}{4008--4016}
  (\bibinfo{year}{2008}).

\bibitem{mathieu_icosahedra}
\bibinfo{author}{Leocmach, M.} \& \bibinfo{author}{Tanaka, H.}
\newblock \bibinfo{title}{Roles of icosahedral and crystal-like order in hard
  spheres glass transition} \bibinfo{pages}{submitted} (\bibinfo{year}{2012}).

\bibitem{AuerR}
\bibinfo{author}{Auer, S.} \& \bibinfo{author}{Frenkel, D.}
\newblock \bibinfo{title}{Numerical simulation of crystal nucleation in
  colloids}.
\newblock \emph{\bibinfo{journal}{Adv. Polym. Sci.}}
  \textbf{\bibinfo{volume}{173}}, \bibinfo{pages}{149--207}
  (\bibinfo{year}{2005}).

\bibitem{filion}
\bibinfo{author}{Filion, L.}, \bibinfo{author}{Hermes, M.},
  \bibinfo{author}{Ni, R.} \& \bibinfo{author}{Dijkstra, M.}
\newblock \bibinfo{title}{Crystal nucleation of hard spheres using molecular
  dynamics, umbrella sampling, and forward flux sampling: A comparison of
  simulation techniques}.
\newblock \emph{\bibinfo{journal}{J. Chem. Phys.}}
  \textbf{\bibinfo{volume}{133}}, \bibinfo{pages}{244115}
  (\bibinfo{year}{2010}).

\bibitem{russo_gcm}
\bibinfo{author}{Russo, J.} \& \bibinfo{author}{Tanaka, H.}
\newblock \bibinfo{title}{Selection mechanism of polymorphs in the crystal
  nucleation of the Gaussian core model}.
\newblock \emph{\bibinfo{journal}{Soft Matter}} \textbf{\bibinfo{volume}{8}},
  \bibinfo{pages}{4206--4215} (\bibinfo{year}{2012}).

\bibitem{likos}
\bibinfo{author}{Likos, C.~N.}
\newblock \bibinfo{title}{Effective interactions in soft condensed matter
  physics}.
\newblock \emph{\bibinfo{journal}{Phys. Rep.}} \textbf{\bibinfo{volume}{348}},
  \bibinfo{pages}{267--439} (\bibinfo{year}{2001}).

\bibitem{pristipino_pre}
\bibinfo{author}{Prestipino, S.}, \bibinfo{author}{Saija, F.} \&
  \bibinfo{author}{Giaquinta, P.~V.}
\newblock \bibinfo{title}{Phase diagram of the Gaussian-core model}.
\newblock \emph{\bibinfo{journal}{Phys. Rev. E}} \textbf{\bibinfo{volume}{71}},
  \bibinfo{pages}{050102} (\bibinfo{year}{2005}).

\bibitem{wedekind}
\bibinfo{author}{{Wedekind}, J.}, \bibinfo{author}{{Strey}, R.} \&
  \bibinfo{author}{{Reguera}, D.}
\newblock \bibinfo{title}{{New method to analyze simulations of activated
  processes}}.
\newblock \emph{\bibinfo{journal}{J. Chem. Phys.}}
  \textbf{\bibinfo{volume}{126}}, \bibinfo{pages}{134103}
  (\bibinfo{year}{2007}).

\bibitem{moore2010freezing}
\bibinfo{author}{Moore, E.~B.}, \bibinfo{author}{de~La~Llave, E.},
  \bibinfo{author}{Welke, K.}, \bibinfo{author}{Scherlis, D.} \&
  \bibinfo{author}{Molinero, V.}
\newblock \bibinfo{title}{Freezing, melting and structure of ice in a
  hydrophilic nanopore}.
\newblock \emph{\bibinfo{journal}{Phys. Chem. Chem. Phys.}}
  \textbf{\bibinfo{volume}{12}}, \bibinfo{pages}{4124--4134}
  (\bibinfo{year}{2010}).

\bibitem{romano_tetrahedral}
\bibinfo{author}{Romano, F.}, \bibinfo{author}{Sanz, E.} \&
  \bibinfo{author}{Sciortino, F.}
\newblock \bibinfo{title}{Crystallization of tetrahedral patchy particles in
  silico}.
\newblock \emph{\bibinfo{journal}{J. Chem. Phys.}}
  \textbf{\bibinfo{volume}{134}}, \bibinfo{pages}{174502}
  (\bibinfo{year}{2011}).

\bibitem{reinhardt2011free}
\bibinfo{author}{Reinhardt, A.} \& \bibinfo{author}{Doye, J. P.~K.}
\newblock \bibinfo{title}{Free energy landscapes for homogeneous nucleation of
  ice for a monatomic water model}.
\newblock \emph{\bibinfo{journal}{J. Chem. Phys.}}
  \textbf{\bibinfo{volume}{136}}, \bibinfo{pages}{054501}
  (\bibinfo{year}{2012}).

\bibitem{moore2010ice}
\bibinfo{author}{Moore, E.~B.} \& \bibinfo{author}{Molinero, V.}
\newblock \bibinfo{title}{Ice crystallization in waterâs âno-manâs
  landâ}.
\newblock \emph{\bibinfo{journal}{J. Chem. Phys.}}
  \textbf{\bibinfo{volume}{132}}, \bibinfo{pages}{244504}
  (\bibinfo{year}{2010}).

\bibitem{moore2011cubic}
\bibinfo{author}{Moore, E.} \& \bibinfo{author}{Molinero, V.}
\newblock \bibinfo{title}{Is it cubic? Ice crystallization from deeply
  supercooled water}.
\newblock \emph{\bibinfo{journal}{Phys. Chem. Chem. Phys.}}
  \textbf{\bibinfo{volume}{13}}, \bibinfo{pages}{20008} (\bibinfo{year}{2011}).

\end{thebibliography}


\end{document}